%% file: main.tex
\documentclass[conference]{IEEEtran}
\IEEEoverridecommandlockouts
\usepackage{cite}
\usepackage{fancyhdr}
\usepackage{amsmath,amssymb,amsfonts}
\usepackage{algorithmic}
\usepackage{graphicx}
\usepackage{textcomp}
\usepackage{xcolor}
\usepackage{url}

\def\BibTeX{{\rm B\kern-.05em{\sc i\kern-.025em b}\kern-.08em
    T\kern-.1667em\lower.7ex\hbox{E}\kern-.125em}}

\fancypagestyle{FirstPage}{
\lfoot{978-1-6654-8045-1/22/\$31.00 ©2022 IEEE} 
}

\begin{document}

\title{DCPViz: A Visual Analytics Approach for \\ Downscaled Climate Projections}

\def\plainauthor{\IEEEauthorblockN{Abdullah-Al-Raihan Nayeem $^{1}$, Huikyo Lee$^{2}$, Dongyun Han$^{3}$, \\ Mohammad Elshambakey$^{4}$, William J. Tolone$^{1}$, Todd Dobbs$^{1}$, Daniel Crichton$^{2}$, and Isaac Cho$^{1,3}$}}
\author{\plainauthor\\ \IEEEauthorblockA{
{$^{1}$College of Computing and Informatics, University of North Carolina at Charlotte, Charlotte, United States} \\ 
{$^{2}$Jet Propulsion Laboratory, California Institute of Technology, Pasadena, United States} \\
{$^{3}$Computer Science, Utah State University, Logan, United States}\\
{$^{4}$City of Scientific Research and Technological Applications, Alexandria, Egypt}\\
}}

\maketitle


\begin{abstract}
This paper introduces a novel visual analytics approach, DCPViz, to enable climate scientists to explore massive climate data interactively without requiring the upfront movement of massive data. Thus, climate scientists are afforded more effective approaches to support the identification of potential trends and patterns in climate projections and their subsequent impacts. We designed the DCPViz pipeline to fetch and extract NEX-DCP30 data with minimal data transfer from their public sources. We implemented DCPViz to demonstrate its scalability and scientific value and to evaluate its utility under three use cases based on different models and through domain expert feedback.
\end{abstract}


\section{Introduction}
\begin{figure*}[t]
  \centering
    \includegraphics[width=0.95\linewidth]{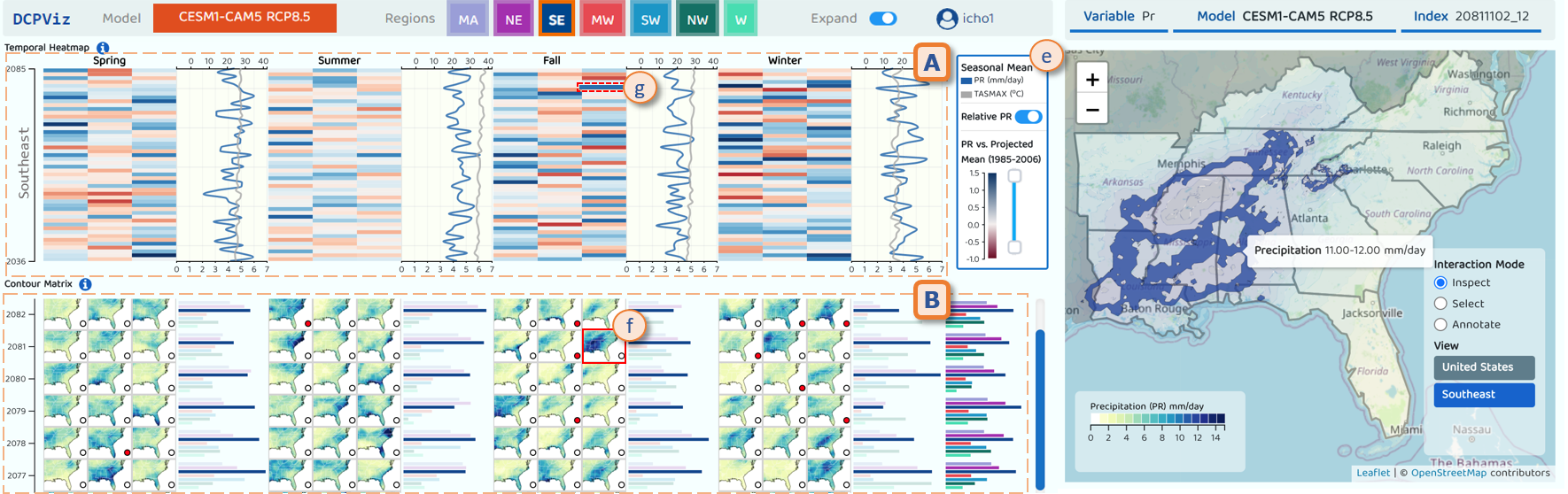}
  \caption{
  DCPViz employs coordinated multi views (CMV) to explore high-resolution statistically downscaled climate projections. It includes a regional view of (A) \textit{Temporal Heatmap} - presents relative monthly-averaged projection of the climate variable (e.g., precipitation)
  (B) \textit{Contour Matrix} - aligned with temporal heat map illustrates geo-coordinated projections, and (C) \textit{Map view} - renders multi-layered interactive map for exploring geospatial contours.
  }
  \label{fig:teaser}

\end{figure*}

\thispagestyle{FirstPage}
\input{sections/1-introduction}

\input{sections/2-related-works}
\input{sections/3-background}
\input{sections/4-va-system}
\input{sections/5-evaluation}
\input{sections/6-discussion}
\input{sections/7-conclusion}

\section*{Acknowledgment}
This paper was supported by the US National Science Foundations (NSF) Data Infrastructure Building Blocks
(DIBBs) Program (Award \#1640818).


\bibliographystyle{IEEEtran}
\bibliography{main}

\end{document}

%% file: sections/1-introduction.tex

%

In the United States (U.S.), the National Climate Assessment (NCA) \cite{Jacobs2016} is the nation’s leading resource for exploring and addressing adaptation and mitigation questions related to climate change across the various sectors of society. To explore and assess the potential impacts of future climates, the NCA uses statistically downscaled climate projections at a high spatial resolution of 10km or finer. Statistical downscaling (SD) is a correction technique for mitigating biases in important variables of interest (e.g., temperature and precipitation). This technique uses observations to estimate the bias-corrected variables appropriately and at a higher spatial resolution than the original model outputs.

Recognizing the importance and need for bias-corrected statistical downscaling to make reliable future climate projections (e.g., \cite{Wootten2021}), NASA’s Earth eXchange project released the Downscaled Climate Projections (NEX-DCP30) at 800m resolution to support climate change research \cite{Nemani2015, Alder2015}. The NEX-DCP30 provides climate projection data, including monthly averaged precipitation (\textit{pr}), daily maximum temperature (\textit{tasmax}), and daily minimum temperature (\textit{tasmin}), for the contiguous U.S. for the 115 years between 1985 and 2099. The high-resolution spatiotemporal dataset is organized according to the Network Common Data Form (NetCDF) format \cite{edward2008experience} and publicly available on the Amazon Simple Storage Service (S3) \cite{nexdcp30}.

There are, however, many challenges that limit the utility of the NEX-DCP30 and similar downscaled climate projection (DCP) data for climate research and relevant scientific analyses. First, due to the size of such data, it is often problematic for climate scientists to migrate data to local storage, which is often necessary for data analysis and exploration. For example, each NetCDF file contains 5 years of data over the contiguous U.S. at 800m resolution, which makes each file size approximately 2GB (or, 17TB for the complete archive) \cite{Lee2019, Nemani2015}. Similarly, CMIP6 \cite{Eyring2016} is expected to produce more than 30 PB of simulation results through the Earth System Grid Federation. Second, existing tools for visualizing, analyzing, and exploring data like the NEX-DCP30 (e.g., Panoply \cite{2018AGUFMIN21B35A}, NCCV \cite{Alder2015}) are limited. Existing tools normally depict only a few static plots with minimal support for interactive exploration and comparative analyses. Finally, even with prior knowledge of climate models and their associated data formats (e.g., NetCDF), it is a prohibitive and complex task to convert data into formats that are easily interpretable, understandable, and, thus, useful for decision-makers \cite{Davis2017, Alder2015}. 

In a traditional pipeline, 
analyzing massive climate data 
usually begins with users 
searching and locating high-resolution climate data. Users, then, must manage local storage, prepare computational resources, and request data access. Once access is granted, users must download the (usually massive) data, convert the data into the desired format, analyze the data computationally, and, visualize the converted data for scientific analyses to create, for example, performance benchmarks based on the dataset. Due to the importance of high-resolution climate data to the U.S. NCA \cite{Jacobs2016} and other international climate assessments (e.g., the IPCC) \cite{stockhause2021cmip6}, climate scientists and decision-makers require more innovative and interactive, visual, and cloud-based solutions to overcome the highlighted challenges and maximize the utility of climate data like the NEX-DCP30 to improve our understanding of future climate change at multiple scales (e.g., regional, national).

To address the challenges of analyzing and exploring shared high-resolution climate projection data like the NEX-DCP30, this paper introduces a novel interactive visual analytics approach, called DCPViz, that improves the efficiency and efficacy of climate projection analyses. DCPViz provides data exploration and comparative analyses with better accessibility to shared data, and a streamlined environment for climate scientists and decision-makers to analyze climate projections data. To evaluate our approach, we demonstrated DCPViz to climate scientists, allowed the scientists to explore DCPViz (and associate climate data - i.e., NEX-DCP30), solicited expert feedback, and performed a qualitative evaluation of the system. Our findings revealed that key aspects of DCPViz - its interactivity, accessibility, and sense-making support - assisted the evaluators in climate analyses and improved overall task performance.
The contributions of this paper are:
\begin{itemize}
    \itemsep0em
    \item We provide a novel cloud-based pipeline to extract high-resolution climate data (e.g., NEX-DCP30 data) for an interactive visual analytics system. The pipeline substantially reduces redundant downloads of large-scale data by moving analysis to the site of the data and migrating only the results of the analyses. 
    
    \item We introduce a web-based interactive visual analytics tool, DCPViz, that 
    helps scientists perform analysis, reasoning, and decision-making tasks with spatiotemporal climate projections.
    
    \item We present DCPViz use cases designed and recommended by climate scientists. We also present findings based on expert evaluations. These findings highlight the aspects of DCPViz that assist climate scientists to improve task performance.
\end{itemize}

%% file: sections/2-related-works.tex
\section{Related Work}

To situate the contributions of DCPViz, we summarize related works from two perspectives: i) visualization systems for earth science domains and ii) visualization techniques for relevant data types. Our literature review is further scoped to the challenges of analyzing massive, high-resolution DCP data.

\subsection{Visualization Systems for Earth Science}
Several visualization and modeling platforms have been developed for climate data analysis in recent years. However, to our best knowledge, no previous platforms demonstrate either: i) a visual analytics pipeline with exploratory features for massive, high-resolution DCP data; or ii) sufficient interactive visualizations for exploratory and comparative analyses.


\subsubsection{General Purpose Systems for Earth Science}
Most general-purpose visualization systems demand data to be organized according to a well-defined format prior to visualizing, exploring, and analyzing the data. In addition, visualization systems until recently were not compatible with cloud-based architectures.
Analytics systems for large-scale data such as Community Data Analysis Tool (CDAT) \cite{potter2009visualization, santos2013uv}, Climate Engine \cite{Huntington2017}, MATLAB \cite{hanselman1996mastering}, and NCAR \cite{NCAR2019}, and GrADS \cite{Berman2001} have been widely applied to the earth science domains. The goal of these platforms is to provide data mapping and time-series visualization features to users of geospatial data. Despite comprehensive analytical features, these systems do not offer integrated visualizations for analyzing spatiotemporal patterns.

\subsubsection{Visualization Systems for Climate Change}
Climate scientists work with diversified data that frequently demand visualizations and interactions specifically tailored for the analysis of massive, high-resolution spatiotemporal data. Hence, most of the special purpose systems have domain experts involved in the design, development, and evaluation process \cite{nockevisclimate} according to task-specific requirements such as operational weather forecasting, climate change assessments, etc.



HI-RAMA \cite{McLean2020} and VAPOR \cite{Li2019} are domain-specific systems to help climate scientists predict and explore the impact of potential climate change.
NCCV \cite{Alder2015} offers exploratory visualizations for NEX-DCP30 temperature and precipitation data. To our best knowledge, NCCV is the only tool focused on NEX-DCP models, but it was developed based on Flash player which is no longer available. Similar visualization systems have been developed to facilitate deep exploration and comparative analysis of climate projections \cite{Blower2013, Herring2017, Sharma2018, Li2019}. But these systems limit analytical support to specific variables or models. 
NASA developed an online analysis platform, Giovanni \cite{2007EOSTr8814A}, that limits its study of geophysical variables including precipitation and temperature, as well, restricts users to load preferred data and processing scripts.

Several platforms have been developed (e.g., MeteoInfo \cite{ramirez2007meteoinfo}, WebGlobe \cite{Sharma2018}, Panoply \cite{2018AGUFMIN21B35A}) 
to analyze the data encoded in the NetCDF format, an abstraction for storing multidimensional data, and a popular data encoding format for climate variables \cite{Rew1990, Sharma2018, Li2019}. 
These tools are sufficient to generate a geospatial view from NetCDF data but lack interactive features for exploring spatial patterns.


\subsection{Interactive Visualization Techniques}
Climate variables are often attributed to geolocation and timestamp information. As such, climate model evaluations utilize geospatial visualizations and time charts \cite{nockevisclimate}. Geospatial visualizations commonly enable data exploration based on geographic location \cite{Cartwright2000, Alder2015}. Time charts are often used to identify trends over time. Information access, navigation, user interaction, and manipulation are some common affordances of interactive geospatial visualization systems. While visualization has a lengthy history with climatology, researchers continue to search for better visual approaches to explore complex environmental information in a manner that is both understandable and useful for a broader range of tasks \cite{Cartwright2000}. As such, visual analytics systems often leverage other visual artifacts to support geospatial visualization. For example, area maps are used to plot the area of modeled impact \cite{2007EOSTr8814A, Li2019}. 
Parallel coordinates are utilized to analyze the association in multivariate climate data \cite{poco2014}.
Histograms, bar plots, and hierarchical ensemble clustering are popularly leveraged in several systems to compare the climate models in different spatiotemporal granularity \cite{ kappe2019analysis, nayeem2021visual}. 
However, these tools and systems provide insufficient support for coordinated multi-visualizations with interlinked user interactions, which are very useful for pattern detection and comparative analysis.

DCPViz, however, allows climate scientists to explore spatiotemporal patterns and covariability among projected climate variables using interactive spatiotemporal and  geospatial visualization. The interactive features also allow the user to interact with its system pipeline to transform the data and link them to coordinated visualizations.



%% file: sections/3-background.tex
\begin{figure*}[t]
    \centering
    \includegraphics[width=0.9\linewidth]{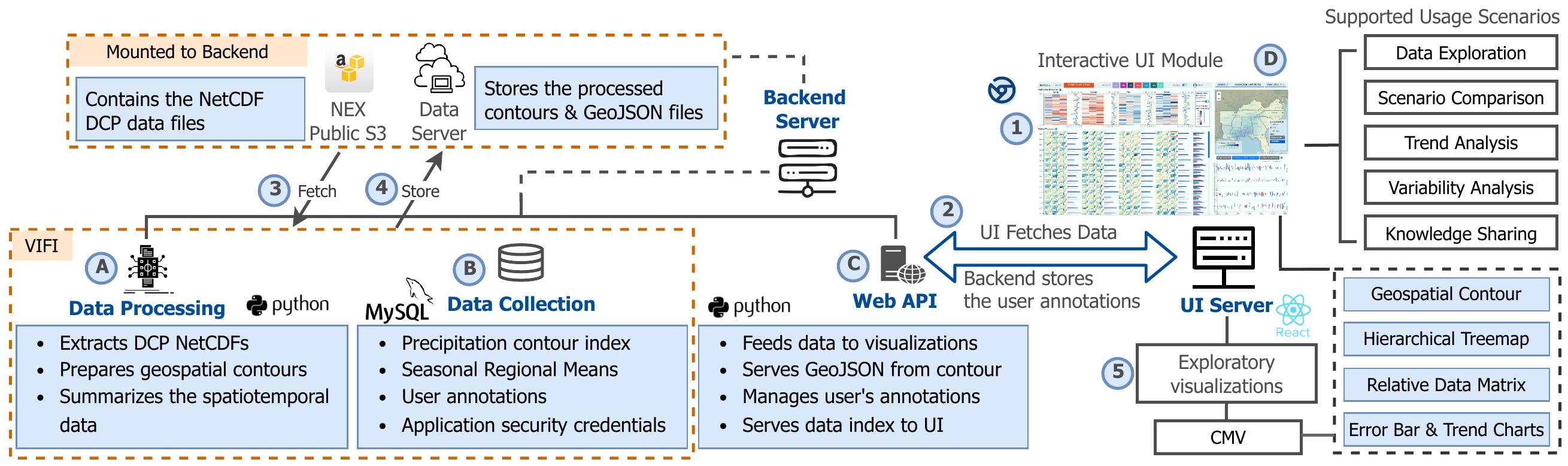}
    \caption{The proposed DCPViz pipeline incorporates 3 distributed servers for exploring spatiotemporal climate projections: 1) Data Server, 2) Backend Server, and 3) UI server. There are four major modules in our pipeline: A) \textit{Data Processing Module} - performs transformation and analysis in a cloud environment (VIFI), B) \textit{Data Collection Module} - stores the resulting indexed data, C) \textit{Web API Module} - maintains the transactions between data and UI servers, and D) \textit{Interactive UI} - a web-based module to allow users simultaneous access to the system.}
    \label{fig:system-pipeline}
\end{figure*}

\section{Design Method \& Requirements}
To design DCPViz, our team of visualization researchers and climate scientists met weekly for more than one year to define design requirements, discuss climate science workflows, and conduct several research and design interactions. 
In addition, we reviewed related works and the goals of NEX for the NCA \cite{Nemani2015, 2016AGUFMGC31G1182N}. These goals include: a) engaging scientists in addressing climate change, b) improving the scope of climate research and analysis, and c) facilitating better accessibility to data and software tools to encourage collaboration and knowledge sharing.
Based on the literature, we identified five primary design requirements:

\begin{enumerate}
    \item
    [R1]
    \textbf{To facilitate data preparation without the overhead of massive data downloads.} 
    The system should extract spatiotemporal climate projections from NetCDF data files stored in a cloud environment efficiently and eliminate redundant downloads of large data files in order to run exploratory analyses.
    
    \item
    [R2]
    \textbf{To enable interactive transformation and filtering to multidimensional data.}
    The system should provide seamless accessibility to interactive features for data transformation and filtering for exploratory and sense-making tasks. Leveraging a web-based protocol for data access will also support a NEX objective by enabling better data accessibility to climate scientists.
    
    \item
    [R3]
    \textbf{To enable climate scientists to perform visual exploratory analyses.}
    The system should offer a web-based interactive interface that allows climate scientists to access model projections for the emission scenarios. The interface should provide climate scientists visual components to perform analysis tasks, reason over analytical results, and identify latent insights.
    
    \item
    [R4]
    \textbf{To facilitate trend and covariance analysis in different spatiotemporal granularity.}
    The system should offer interactive visualizations to observe underlying patterns and identify covariance among climate variables at multiple spatial (e.g., region, state, county) and temporal (e.g., annual, seasonal, monthly) granularities.
    
    \item
    [R5]
    \textbf{To enhance collaboration and sharing among climate scientists.}
    The system should serve as a platform for climate scientists where they can perform data exploration and analyze climate variables simultaneously. The system should facilitate further collaboration by enabling the sharing of key findings among climate scientists.

\end{enumerate}

To satisfy the aforementioned requirements, we designed a novel visual analytics pipeline for DCPViz. This pipeline contains four modules: data processing, data collection, (RESTful) web API, and the UI, which provides interactive visual components. Fig. \ref{fig:system-pipeline} illustrates the modules and process flows. 

    \textbf{The Data Processing Module} enables data transformation and analysis over distributed data by leveraging VIFI   \cite{vifiA, vifiB}. This module runs on a remote server to access the NetCDF files from a  mounted S3 bucket containing the DCP data (Fig. \ref{fig:system-pipeline}A and \ref{fig:system-pipeline}-3). These data are extracted from the cloud in an efficient matter to facilitate data preparation and reduce the overhead of massive data downloads prior to analysis \textbf{(R1)}.
    
    \textbf{The Data Collection Module} stores the extracted data in a shared repository and obtains the data index to facilitate accessibility `on demand' (Fig. \ref{fig:system-pipeline}B and \ref{fig:system-pipeline}-4). 
    Each index is composed of a unique code for the model name, variable name, temporal identifier, and the DCP dataset name (e.g., \textit{`NEX-DCP\_CESM1-CAM5\_pr\_2021-03-01'}). Each index identifies its original NetCDF file.
    The shared repository stores the indices for the entire dataset (e.g., spatiotemporal precipitation, maximum and minimum temperatures). When the user focuses on a certain segment in the timeline or a spatial region, the interface passes a request to the API module to fetch 
    the required data \textbf{(R2)}.

    \textbf{The Web API Module} is a backend server that extracts a subset of the data from the remote site using specified filters (Fig. \ref{fig:system-pipeline}C). To eliminate the need for prerequisite knowledge about the climate model and data format, the system automatically pre-processes the data into the requested format. This is a standalone service that provides precipitation and temperature data to the UI module and associated analysis scripts \textbf{(R2)}.
    
    \textbf{The UI Module} provides Coordinated Multi Views (CMV) to allow users to investigate climate change on a regional scale leveraging high-resolution NEX-DCP30 data \cite{nockevisclimate}. This module provides not only conventional plots according to the traditional workflow 
    but also an interactive visual interface (Fig. \ref{fig:system-pipeline}D). The interface supports exploratory analyses on future climate change over spatial and temporal domains \textbf{(R3)}. 
    Moreover, the interface displays climate variability and a summary of overall seasonal changes between \textit{pr} and \textit{tasmax} and across the 7 NCA regions in the contiguous U.S. \textbf{(R4)}. Furthermore, DCPViz supports textual annotations in the map and matrix views to  
    share key findings with other users on DCPViz \textbf{(R5)}. Section \ref{sec:user-interface} describes the UI in detail.
    

We integrated these modules to build our proposed visual analytics system. The UI module interprets a sequence of events from the user and delegates them to the distributed modules to support requested functionalities. 

%% file: sections/4-va-system.tex
\section{Visual Analytics Approach}
In this section, we describe DCPViz and illustrate how it meets our design requirements. 

\subsection{The NEX-DCP30 Dataset}
NEX is a collaborative platform to facilitate the analysis and forecast of climate data \cite{Nemani2015}. NEX projections provide several statistically downscaled datasets via a public S3 bucket. Among these datasets, NEX-DCP30 is the dataset with the highest spatial resolution of about 800m. Three variables are included in NEX-DCP30 (precipitation - \textbf{\textit{pr} (\(mm/day\))}, daily maximum temperature - \textbf{\textit{tasmax}} (K), and daily minimum temperature - \textbf{\textit{tasmin}} (K) later converted to (\(^oC\))). They are derived from 36 climate models and observational data. We use two of these models for DCPViz: Community Atmospheric Model version 5 (CESM1-CAM5 \cite{Meehl2013}) and Goddard Institute for Space Studies Model E2 coupled with the Russell ocean model (GISS-E2-R \cite{kelley2020giss}). NEX-DCP30 contains 115 years of monthly data for the contiguous U.S. 
Data from 1985 to 2005 are historical and data from 2006 to 2099 are projected under three RCP scenarios (2.6, 4.5, and 8.5). 



\subsection{Data Transformation \& Analysis}\label{sec:data-analysis}
\begin{figure}[t]
    \centering
    \includegraphics[width=0.9\linewidth]{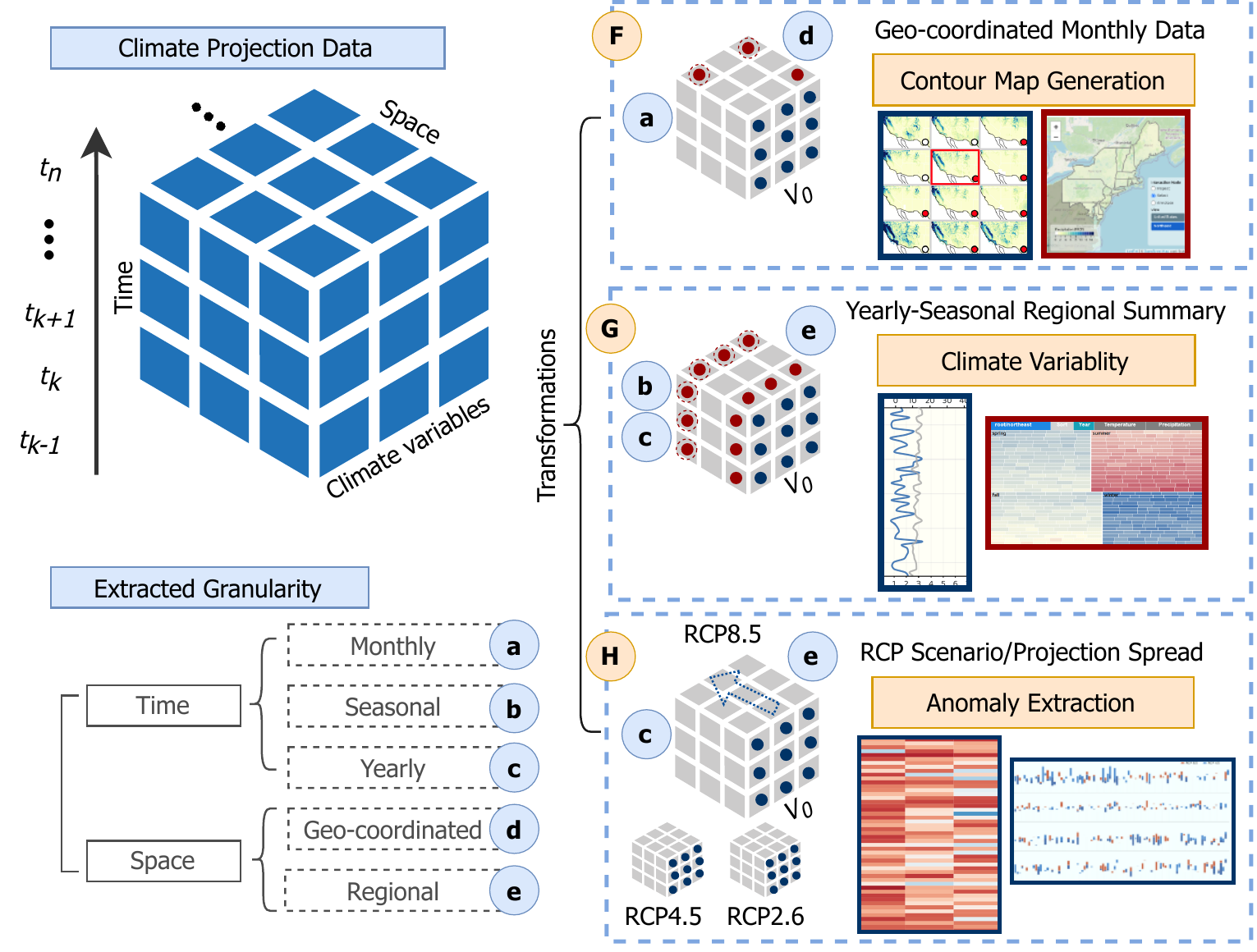}
    \setlength{\belowcaptionskip}{-12pt}
    \caption{Spatiotemporal transformations of the projection data for exploratory analysis. The transformations illustrate the extracted hierarchical levels of spatial and temporal granularity.}
    \label{fig:data-transformation}
\end{figure}

We leverage VIFI, an open-source tool that enables the analysis of distributed fragmented data. VIFI enables data sharing by migrating analytics (often lightweight) to data locations rather than by migrating massive data \cite{vifiA, vifiB}. 
Users can perform a variety of analyses, specified as analytical workflows (Fig. \ref{fig:data-transformation}), over distributed data stored at multiple locations.
VIFI workflows are implemented for data-driven discoveries from distributed climate data \cite{vifiA}.
We utilize VIFI to enable contour map generation, analyze climate variability and identify anomalous patterns from the RCP scenario projections (Fig \ref{fig:system-pipeline}A-B). 
Workflow execution begins with query generation based on user-selected parameters such as model name, climate variables, experiment ID, and year range. Once the backend server accesses the data corresponding to specified parameters, the Data Processing Module (Fig. \ref{fig:system-pipeline}A) triggers the VIFI workflows to perform data extraction, transformation, and analysis.
To demonstrate this process in DCPViz, we provide an illustration using \textit{pr} and \textit{tasmax} in the NEX-DCP30 models.

\textbf{Contour Map Generation.} 
We extract NetCDF files using VIFI data orchestration to retrieve geo-coordinated monthly averaged climate projections (e.g, \textit{pr} and \textit{tasmax}). The workflow extracts 60 projected snapshots (one for each month) from each NetCDF file that recorded the spatiotemporal climate variables. Next, the contour generation process (Fig. \ref{fig:data-transformation}F) transforms the geo-coordinated data to the GeoJSON format.
While extracting the data remotely, the collection module attaches an index to the generated GeoJSON using the search parameters.

\textbf{Spatiotemporal Granularity Extraction.}
We apply a series of transformations to geo-coordinated monthly projection data to extract the spatiotemporal granularity (Fig. \ref{fig:data-transformation}).
We apply an NCA’s region mask on the projected data to segment the temporal variable intensity into seven regions (Northeast, Southeast, Midwest, Southwest, Northwest, Northern Great Plains, and Southern Great Plains).\footnote{The 4th NCA report segments U.S. in 10 regions \cite{reidmiller2017fourth} but Alaskan, Hawaiian, and U.S. Caribbean are excluded in the NEX-DCP30 data} Then, we utilized the segmented data to calculate seasonal and yearly regional means for the projected variables. These data are further exploited to analyze and assess climate variability in the projected models.

\textbf{Extracting Anomalous Projections.}
Lastly, we derive measures to extract anomalous patterns from the climate projections. A previously calculated yearly mean is leveraged to measure the percentage change between the yearly mean and the selected decades of retrospective mean (e.g., 1985-2006), what we call \textit{`Relative Intensity'}. Seasonal regional retrospective means are described as \( RM(t, s, r) \) in the following:
\begin{equation}\label{eq:retro-mean}
RM(t, s, r) = \frac{1}{3 \Delta t} * \sum_{y=t_{0}}^{t_{1}} \sum_{m=s}^{s + 3} X_{r} (y, m)
\end{equation}
where \(t\) represents a year range within retrospective timeline, \(s\) represents the season, and \(r\) represents the region. \(X_{r} (y,m)\) provides the monthly-averaged regional mean for the climate variables. \( RM(t, s) \) is leveraged in calculating the \textit{`Relative Intensity'}, \(RI (t, y, m, r)\) for \textit{pr} described as follows:
\begin{equation}
RI (t, y, m, r) = \frac{\mid RM(t, \lfloor m/3 \rfloor, r) - X_{r}(y, m) \mid}{RM(t, \lfloor m/3 \rfloor, r)}
\end{equation}
The intensity fluctuation is often projected within a wide range of variations, which can overlook anomalies at certain data points. For example, a 0.5 \(mm/day\) change in \textit{pr} may not be noticeable in a snapshot compared to the overall range. However, this small change could be significant if the \textit{pr} trends around 2 \(mm/day\) in that month/season.
As such, \(RI (t, y, m, r)\) can amplify anomalous patterns. 

The \textbf{Data Collection Module} (Fig. \ref{fig:system-pipeline}B) maintains a repository of extracted and transformed data as well as analytical results. 
We store these data in a manner that allows for efficient filtering while exploring projections from the entire timeline. 
In addition, this module manages the user’s credentials and annotations. 

The \textbf{Web API Module} transforms collected data as per the required structure for the interactive visualizations before responding to the requests from the UI Module. All of these processes are executed remotely to limit data migration and the need for user knowledge of data format and organization. 

\input{sections/user-interface}


%% file: sections/user-interface.tex
\begin{figure*}[t]
  \centering
    \includegraphics[width=0.9\linewidth]{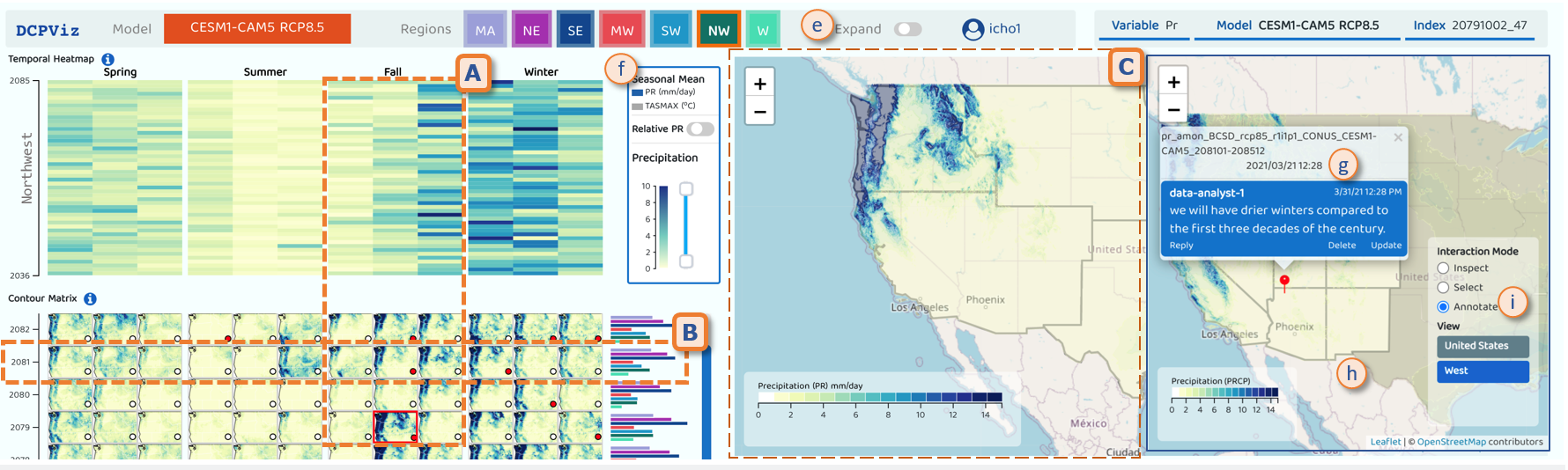}
  \caption{Overview of the DCPViz interface. (A) Temporal heatmap mapped with contour matrix presents a seasonal progression of geospatial intensity (\textit{pr}) (top to bottom), whereas (B) regional bar chart provides yearly-averaged regional intensity (\textit{pr}) followed by a monthly progression (left to right). (C) Map view enables annotating the geo-coordinated location in addition to interactive depictions of contour matrix snapshots. 
  }
  \label{fig:interface}
\end{figure*}

\subsection{DCPViz: Exploratory Visual Interface}\label{sec:user-interface}
The DCPViz interface presents interactive temporal and spatial visualizations for retrospective and projected climate data. 
The UI leverages CMV to enable 
rich user interaction. It consists of three main views: 1) a map view (Fig. \ref{fig:interface}C), 2) an adaptive spatiotemporal view (Fig. \ref{fig:interface}A-B), and 3) a reconfigurable summary panel (Fig. \ref{fig:summary-time-series}, \ref{fig:summary-rcp-bar}, and \ref{fig:summary-tree-map}). We implemented the interface for exploring the projection of historical \textit{pr} and \textit{tasmax} data across the U.S.. The interface is developed using open-source web libraries.\footnote {The interface is publicly available at  \url{https://esva.jpllab.net/}} 

\subsubsection{Map View} 

The map view shows a geo-coordinated contour that represents the monthly-averaged intensity across the U.S. regions. The view has three map layers: a base map, a contour map, and an area border map for the contiguous U.S. The base map shows different geographical features of the U.S. such as boundaries, rivers, and highways. 
The geospatial contour map is plotted as a variable intensity layer. The area border map shows the borders of the seven U.S. regions. 

The map view provides three interaction modes (Fig. \ref{fig:interface}i): 1) inspection, 2) selection, and 3) annotation. The inspection mode 
allows users to find areas of similar intensity on the contour layer. Areas 
are grouped by the projected intensity levels, which range from 0-14 mm/day with a yellow-green-blue scheme.
Hovering over a specific location opens a tooltip that shows additional information such as the inspected region and the monthly-averaged value at the selected snapshot. Hovering also strengthens the color intensity to let the user effortlessly differentiate the focused area from other areas (Fig. \ref{fig:teaser}C). The selection mode allows users to indicate attention in a specific location. 
Downscaled climate data are projected at 800m resolution \cite{Lee2019, Nemani2015}. So, concentrating attention on a specific region can provide essential detail that may not be easily observed from the default view. Users can also change the view hierarchy to the entire U.S., a specific region, or a specific state on the map.

The annotation mode allows users to share findings and observations 
on the map view.
Once the user adds an annotation, the view shows a pin to share points of interest with others for further exploration (Fig. \ref{fig:interface}h). 
An additional marker is added to each thumbnail in the spatiotemporal view to indicate whether the annotation is added for that snapshot (Fig. \ref{fig:teaser}B). This interaction also allows users to post annotations directly without marking a specific location on the map.

\subsubsection{Spatiotemporal View}
\label{spatiotemporal-view}
The spatiotemporal view presents thumbnails of  geospatial contours  with the corresponding temporal mean from 2036 to 2099. The entire dataset for the model contains 150 years of data \cite{Lee2019} of which we extracted remotely the most recent 1380 (monthly average for 115 years) recorded/projected snapshots. Observations reveal that it can be difficult to identify trends or mark points of anomaly by traversing snapshots using only geospatial projections on the map. To address this, we developed the spatiotemporal view to render the geospatial data against a temporal axis. We add color-coded legends (Fig. \ref{fig:teaser}f) with corresponding measuring units to denote \textit{pr} and \textit{tasmax} in the visualizations. This view has four visual components (Fig. \ref{fig:teaser}A-B): a temporal heat map, a geo-spatiotemporal contour matrix, a climate variability time series, and a regional bar chart.

\textbf{Temporal Heatmap.} 
The temporal heatmap shows the projected mean \textit{pr} sharing the identical axis with the temporal geospatial contour (Fig. \ref{fig:teaser}C). We used two different measures to set the intensity for the heatmap: 1) mean \textit{pr} for each snapshot (Fig. \ref{fig:interface}A); and, 2) relative \textit{pr} (Fig. \ref{fig:teaser}A) derived as \(RI_{Pr} (t, y, m, r)\) in section \ref{sec:data-analysis}. Fig. \ref{fig:teaser}e provides a control \textit{`Relative PR'} to switch between these measures. We also add the option to filter this view by setting upper and lower limits of \textit{pr}. A similar seasonal grouping is applied in the temporal geospatial thumbnail view. Fig. \ref{fig:interface}A shows an example of the Fall season where the cells in a row represent months in the corresponding year. The temporal heatmap helps users understand the progression of \textit{pr} over time. The yellow-green-blue color scheme is used for monthly-averaged mean in the geospatial visualization \cite{kaye2012mapping}. In the \textit{`Relative PR'} view, we adjust the color scheme to blue-pale yellow-red because it is popular for showing anomalies \cite{kaye2012mapping}, which in our case denotes maximum increase to maximum decrease from blue to red. 

\textbf{Geo-spatiotemporal Contour Matrix.}
When capturing the contour for each geo-coordinated monthly snapshot, we store them as static  images. These images are essential to show a preview to help users identify the point of interest.
DCPViz presents the captured geospatial snapshot against a temporal axis. Therefore, the user obtains both the time information and suggested \textit{pr} intensity before deciding to explore the contour in the map view. We maintain a uniform alignment between the temporal heatmap and geospatial contour matrix to help the user identify the corresponding contours from the heatmap cell. We also group the contour images by seasons (Fig. \ref{fig:teaser}B) to help users perceive the seasonal \textit{pr} trend. Hovering over thumbnails shows enlarged renderings of the contour with metadata for a detailed view.


\textbf{Climate Variability Time-Series.} 
The line charts represent a time series of mean \textit{pr} (blue) and mean \textit{tasmax} (gray) (Fig. \ref{fig:teaser}A).  \textit{pr} and \textit{tasmax} are subset data for the selected region, grouped by season and rendered for each season. Each data point in the line charts presents the yearly seasonal-average \textit{pr} record and their associated change in the \textit{tasmax}.

\textbf{Regional Bar Chart.} The regional bar presents the yearly seasonal-average \textit{pr} grouped by region. The map view provides an indication of the intensity of \textit{pr} in different regions. However, the map view may not be adequate for the user to compare \textit{pr} among the regions. Hence, the regional bar chart view is aligned with the geospatial contour in the temporal matrix, as shown in Fig. \ref{fig:interface}B, to show a precise comparison of \textit{pr} among the regions. In addition, the bar representing the selected region is highlighted to amplify the seasonal comparison among regions when scrolling up or down. A yearly-averaged regional bar chart is also included to reveal yearly changes among the seven regions.


\subsubsection{Summary Views}\label{sec:summary-vis}

\begin{figure}[t]
    \centering
    \includegraphics[width=0.9\linewidth]{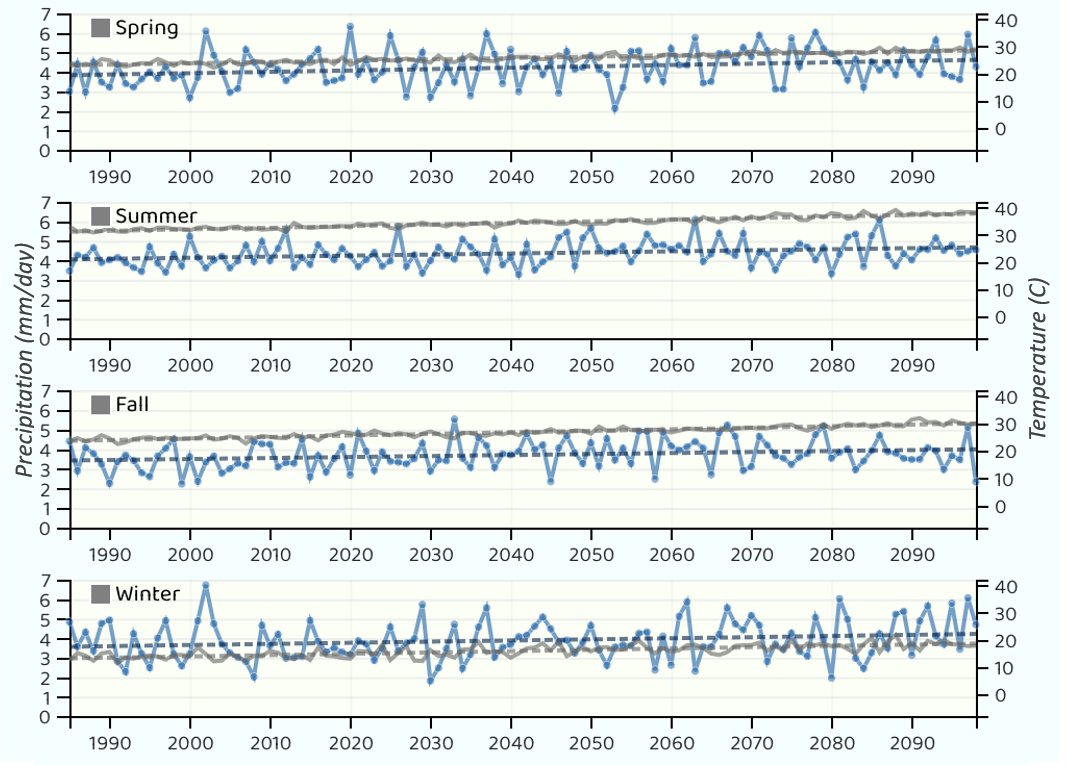}
    \caption{Time-series presents a summary view of yearly-averaged regional climate variability (e.g., projected \textit{pr} and \textit{tasmax} in the southeast region).}
    \label{fig:summary-time-series}
\end{figure}

\begin{figure}[t]
    \centering
    \includegraphics[width=0.9\linewidth]{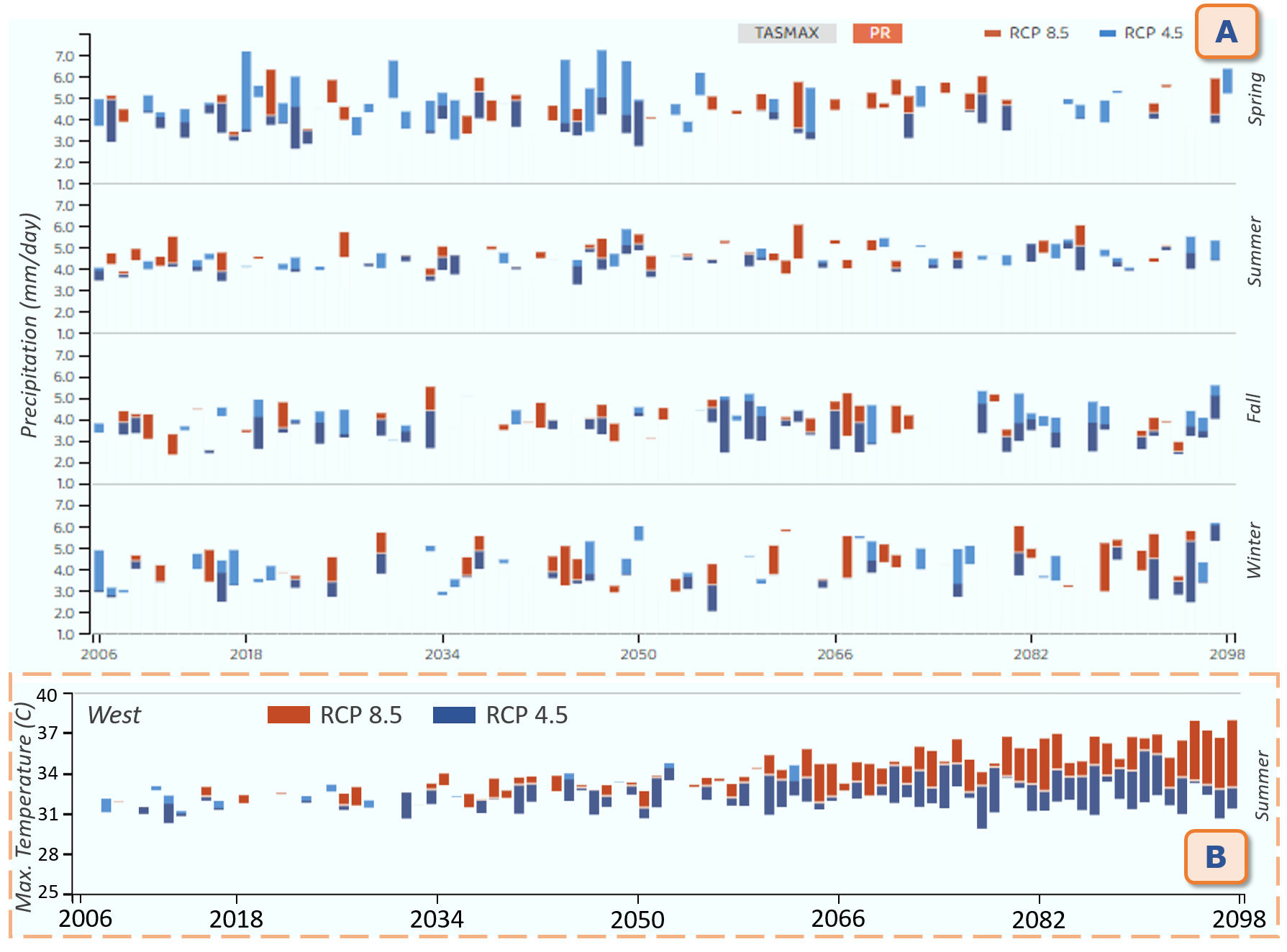}
    \caption{A) Comparative view of RCP scenarios to amplify patterns and anomalies in long-term climate projections. B) Comparing the \textit{tasmax} projections in different RCP scenarios during summer seasons in the western region.}
    \label{fig:summary-rcp-bar}
\end{figure}

\begin{figure}[ht]
    \centering
    \includegraphics[width=0.95\linewidth]{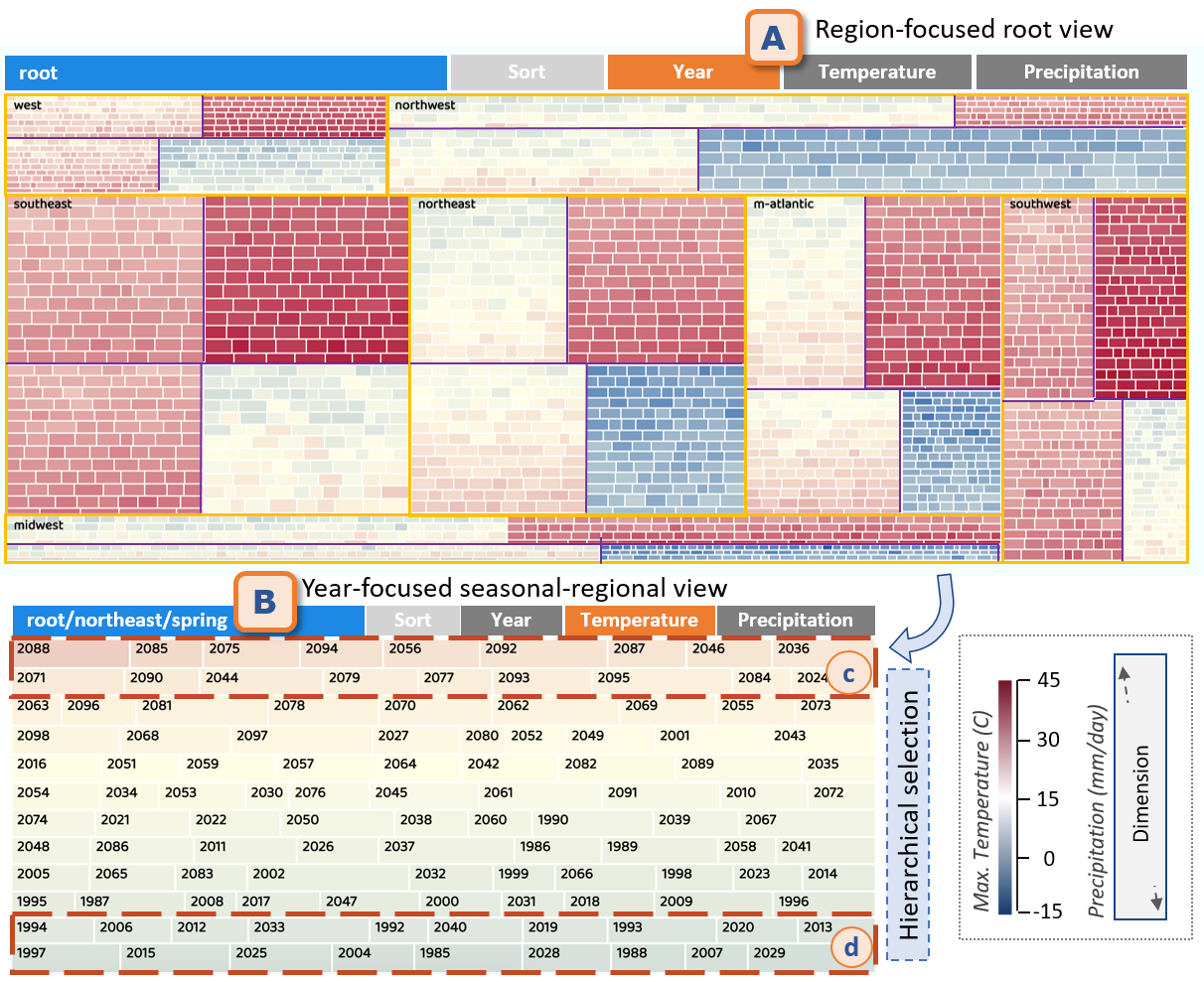}
    \setlength{\belowcaptionskip}{-12pt}
    \caption{The hierarchical treemap supports investigating covariability between \textit{pr} and \textit{tasmax} across the 7 U.S. regions.
    It provides hierarchical selection to reconfigure the focus based on region, season, and year. }
    \label{fig:summary-tree-map}
\end{figure}

Climate projections are often large and complex, and thus difficult to visualize both in totality and with granularity. As such, we provide a reconfigurable summary view to encode and visualize the data. We considered three design factors for the summary view: the visualization purposes, the supporting analysis tasks, and the data type \cite{sarikaya2018design}. We leverage the multi-level spatiotemporal granularity extracted from the projected data during data transformation. The summary view contains three visualizations: a time-series visualization (Fig. \ref{fig:summary-time-series}), an RCP scenario comparison (Fig. \ref{fig:summary-rcp-bar}), and a hierarchical treemap (Fig. \ref{fig:summary-tree-map}A).

\textbf{Time-Series Visualization.} The time-series visualization shows the seasonal \textit{pr} and \textit{tasmax} mean from 1985 to 2098 (Fig. \ref{fig:summary-time-series}). It allows users to explore the trends of projected climate variables, such as \textit{pr} and \textit{tasmax}. It also provides support for direct search to help users find anomalies and patterns in the timeline. We found the time-series visualization to be a suitable candidate for identifying outliers and patterns across a large number of data points \cite{rebbapragada2009finding, nocke2004methods}.


\textbf{RCP Scenario Comparison.} A different design of a stacked bar chart is adopted for visual comparison and anomaly detection across RCP scenario projections (Fig. \ref{fig:summary-rcp-bar}A). 
Each RCP scenario is distinguished by color, where overlapping areas are highlighted with strengthened intensity. 
This visualization presents a yearly-seasonal regional projection spread of the extreme RCP scenarios (e.g., RCP 4.5 and RCP 8.5 as compared to RCP 2.6).
In Fig. \ref{fig:summary-rcp-bar}B, we compare the scenarios for \textit{tasmax} during the summer season in the western region according to CESM1-CAM5. Lower fluctuations among different RCP scenarios are observed in earlier decades in the 21st century. However, \textit{tasmax} is projected to be significantly higher in the later years in the most extreme scenario.

\textbf{Hierarchical Treemap.} 
The treemap visualizes data with a nested structure for two selected climate variables, e.g., \textit{pr} and \textit{tasmax} (Fig. \ref{fig:summary-tree-map}). The treemap is constructed maintaining the following hierarchy of data from the top: region, season, and year (Fig. \ref{fig:data-transformation}).
\textit{pr} is denoted by the dimension of the cells whereas \textit{tasmax} is denoted by the intensity using a blue-yellow-red color scheme for lowest to highest maximums. In the treemap, users can select an area of interest (Fig. \ref{fig:summary-tree-map}B). The treemap presents an overview of the climate co-variability between variables to highlight, for example, which region is getting hotter \& wetter or hotter \& drier under a changing climate.

\begin{figure}[t]
    \centering
    \includegraphics[width=0.7\linewidth]{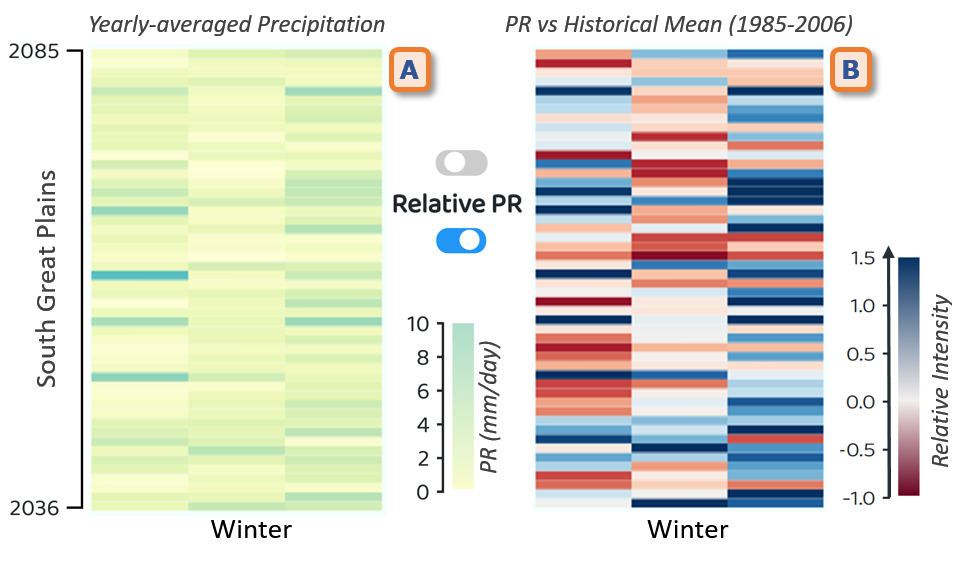}
    \caption{A scenario for the reconfigurable temporal heatmap of the Southern Great Plains in the winter season. The unnoticeable fluctuations in A) yearly-averaged intensity are strengthened in the (B) relative view by plotting the projections against a seasonal historical mean.}
    \label{fig:heatmap_onRelativePR}
\end{figure}



%% file: sections/5-evaluation.tex
\section{Evaluation}
Climate scientists provided a qualitative evaluation of DCPViz to determine its effectiveness at meeting its design requirements \textbf{(R1-R5)}, characterized the  scientific value of climate data analyses, and demonstrated our research contributions. In this section, we describe the DCPViz evaluation metrics, the case studies utilized, and domain expert feedback.

\subsection{Qualitative Evaluation Metrics}
Evaluations of climate science visualizations span numerous qualitative and quantitative dimensions. For our evaluation, we focused on qualitative metrics that assess DCPViz based on its ability to reveal nuances of climate change, intricately visualize massive spatiotemporal data, and depict multivariate association factors for climate change \cite{johansson2010evaluating}. 
To assess the dimensions of interlinked analysis tasks, communication, and decision-making in climate research, we used evaluation metrics from the following categories: \textbf{Visualization} - ability to provide inexplicable insights that reduce time and complexity of analysis, \textbf{Interaction} - ability to coordinate multiple views and exclude irrelevant parameters and values from the view, and \textbf{Presentation} - elements utilized such as visual components, colors, legends, and tooltips to present the data demonstrates self-exploratory capabilities.

\subsection{Evaluation Procedure}
We employed a survey to solicit feedback from domain experts on DCPViz.
Seven climate scientists, who are experts in analyzing NEX-DCP30 data, participated in the survey. 
The survey contained three sections. First, participants were asked to complete the pre-questionnaire regarding their experience with climate data, analysis tasks, and the tools and data analysis libraries they utilize in their analysis. Next, participants watched a video that demonstrated DCPViz functions. 
Then, they interacted with the UI to explore NEX-DCP30 data. Finally, participants were asked to complete a post-questionnaire that inquires about: 
1) Preference of the visualizations, 2) Features that support their analysis tasks, 3) Performance of the CMVs for 
explanatory capabilities and features. 

\subsection{Use Cases}
We highlight the three use cases where DCPViz demonstrated the greatest value to climate scientists.



\textbf{Comprehensive investigation of the spatiotemporal variability in regional precipitation (UC1).}
The temporal heatmap (Fig. \ref{fig:teaser}A) revealed to domain experts future \textit{pr} changes for each of the four seasons. For example, when the user selects the Southeast U.S. region (Fig. \ref{fig:interface}e) and enables \textit{`Relative PR'} (Fig. \ref{fig:teaser}e), the heat map cells in the winter season are mostly rendered in red colors (Fig. \ref{fig:interface}A). This essentially indicates a relatively drier winter later (2036-2085) compared to the first three decades of the century. The contour matrix 
(Fig. \ref{fig:interface}B) reveals that most of the region's \textit{pr} occurs along the West Coast and Sierra Nevada. Users can move their cursors over the thumbnails to see the spatial patterns in \textit{pr}.
By clicking the thumbnail, the precipitation intensity is displayed on the map view with boundaries of the associated NCA region (Fig. \ref{fig:interface}C). The map view supports zooming to maximize the utility of the high spatial resolution (800m). Making annotations (Fig. \ref{fig:teaser}g) facilitates a comparison of \textit{pr} patterns between dry and wet months. 

\textbf{Analyses of long-term trends in temperature and precipitation under different emission scenarios (UC2).}
The summary view (Fig. \ref{fig:summary-time-series}) reveals in the climate data the overall warming trend of \textit{tasmax} for all the seven NCA regions. The observed warming from 1985 to the present is well known, and the bias-corrected temperature from NEX-DCP30 is expected to represent the observed trend. Unlike the warming trends, however, observed \textit{pr} trends in the climate data, as revealed by the relevant visualizations, appear not to be significant across the contiguous U.S..
By default, the line and bar charts represent CESM1-CAM5 model predictions under the RCP 8.5 emissions scenario, which represents high emissions of greenhouse gases without any mitigation efforts. Using DCPViz, domain experts were able to observe that across all seven NCA regions and four seasons, the associated warming trends are higher than those for the RCP 4.5 and 2.6 scenarios. The plots indicate that \textit{pr} changes are more variable than \textit{tasmax} and \textit{tasmin} in all three scenarios. Also, the \textit{pr} predicted by CESM1-CAM5 does not exhibit consistent sensitivity to the emission scenarios across the seven regions (Fig. \ref{fig:summary-rcp-bar}A). 

\textbf{Covariability of seasonal maximum temperature and precipitation (UC3).}
The hierarchical treemap (Fig. \ref{fig:summary-tree-map}) provides scope to observe co-variabilty between the climate variables. 
The cell dimension represents mean \textit{pr} while color intensity scales temperature. The hierarchical treemap also helps domain experts answer key scientific questions related to climate projections - e.g., Are warmer and more humid conditions expected during  the latter decades of the 21st century in the fall season of the Western regions? What about spring in the Northeast region?

\subsection{Domain Expert Feedback}
The domain experts who participated in the survey possessed on average 5 years of experience in climate science research. In the pre-questionnaire, they mentioned that climate data processing and visualization is a moderately to highly difficult task due to the volume and complexity of high-dimensional structure. The domain experts also expressed their desire to have interactive geospatial and time-series views, in contrast to traditional static plots (e.g., by Python or Matlab). 

\textbf{Visualizations.} 
The domain experts mentioned that interactive geospatial visualization (Fig. \ref{fig:interface}C) is worthwhile to have and appreciated the annotation feature as it allows them to share the observations with other scientists. Additionally, inspect user interactions helped them identify the spatial pattern in each snapshot \textbf{(UC1)}. The spatiotemporal view (Fig. \ref{fig:interface}A) was also identified as effective by four experts to understand the temporal progression along with the spatial distribution. The heatmap helped the domain experts quickly inspect the data as the heatmap contains the summary for the whole timeline. The temporal contour matrix aided the domain experts with a quick view of the spatial patterns and their changes over time.
Along with the seasonal and regional time charts, the domain experts identified  the treemap as a less common yet useful visualization for analyzing comparative patterns for multiple variables.

\textbf{Interactions.}
The experts' feedback suggests that the reconfigure and explore interactions are essential for their analysis. One expert stated that the reconfigurable view for summary visualization is convenient for sensemaking long-term patterns from different aggregation perspectives \textbf{(UC2, UC3)}. The exploratory features such as inspecting the map view from the contour thumbnail (Fig. \ref{fig:interface}B) and switching the multi-views based on region selection helped the domain experts to identify quickly interesting facts or insights from the data \textbf{(R3)}. Moreover, the tooltips 
were well appreciated as they provide comprehensive data inspection capabilities and the additional context in the visual analysis.

\textbf{Presentation.}
The domain experts were able to discover their target information from the interactive visualizations while one expert stated the interface became intuitive after the initial learning process.
Moreover, the experts mentioned that the treemap is a good candidate to illustrate the covariability between \textit{pr} and \textit{tasmax} \textbf{(UC3)}. We also learned that the use of information icons, color codes, and legends were deemed useful 
by them.
The domain experts also left us a few suggestions such as - 1) to include annotation for other visualizations (beyond the map), 2) to increase the use of animation in visualizing the projections, and 3) to attempt to simplify the user interface.


Based on the majority of the feedback, we understand DCPViz benefits domain experts as they perform exploratory analyses of climate data  \textbf{(UC1)}. Moreover, the use cases suggest that DCPViz is also useful in examining long-term trends within the NCA regions \textbf{(UC2, UC3)}. 

%% file: sections/6-discussion.tex
\section{Discussion and Limitations}
We proposed a cloud-based visual analytics approach, DCPViz, for making sense of DCP datasets that benefits exploratory analysis by climate scientists. We described specific challenges and identified important design requirements for systems that support the analysis of massive high-resolution spatiotemporal data for climate change research. 
We hypothesized that our proposed requirements would enhance the traditional analysis workflow and support climate scientists in sense-making and decision-making processes. 
We described the four modules that comprise DCPViz (data processing, data collection, web API, and UI) to satisfy the design requirements of the DCPViz pipeline (Fig. \ref{fig:system-pipeline}). In addition, we demonstrated how DCPViz can enhance transitional analysis workflows and support climate scientists' sensemaking and decision-making.  

\textbf{System Scalability.} 
We demonstrated DCPViz with two NEX-DCP30 models (CESMI-CAM5 and GISS-E2-R) in 3 RCP scenarios. The web-based UI provides simultaneous access to climate scientists to visually explore these extracted projection data.
However, climate models are continuously updated to provide high-fidelity simulations by changing parameters. With computational resources becoming more easily accessible, the climate simulations with finer spatial and temporal resolutions (e.g., NEX-GDDP \cite{raghavan2018evaluations}), daily projections are becoming increasingly available.
To incorporate other DCP models and RCP scenarios, we modularized the data processing and collection modules. This allows the scientists to trigger the processing module to extract more model data and store it in the collection module \textbf{(R1)}. 

\textbf{Limitations.} 
To support navigating and exploring the entire model projection, we have added exploratory and explanatory visualizations in the same window. As convenient as it is to gain insight promptly, the interface looks a bit complicated and it requires an initial learning curve to use the interactions. To perform a comparative analysis between projection models and scenarios, the user needs to open two DCPViz interfaces with the participant models or scenarios. In the future, we aim to support visual analysis of multi-model comparisons in DCPViz.

DCPViz enables the exploration of large volumes of data by creating  visual components. We learned from the user study that system performance is affected by client-side network status and computational capacity. Despite the clear benefits of climate science analyses, system performance improvement will be part of our future work. Implementing combinable tabs \cite{jiang2009combinable} and adopting the progressive visualization pipeline \cite{7563865} can ease users' analytical vision while the 
processing is running on background.
Employing a server-side tiled map technique 
for the map view can reduce the rendering time for geospatial contours to enhance the overall performance.

%% file: sections/7-conclusion.tex
\section{Conclusion and Future Work}
In this paper, we introduced DCPViz, a novel visual analytics approach to analyze and explore NEX-DCP30 data. 
We presented three use cases provided by climate scientists while interacting with the DCPViz interface. Experts' feedback is utilized to evaluate the proposed visual analytics approach. While the feedback we received from the experts is promising, we identified several future directions for further research. Our future work will encompass a data mining architecture that is fully integrated with a distributed framework to optimize the data transformation and analysis time. In addition, we will utilize the extracted data from multiple DCP models to assess the climate projection uncertainty. To evaluate the model projections for each climate variable, we plan to add observational datasets, more analytical workflows, comparative visualizations, and interactive features on the interface. Finally, we plan to conduct a more extensive user evaluation by gathering climate scientists in focus groups to evaluate in greater detail the usability and efficiency of DCPViz.
